\documentclass{aa} %https://www.aanda.org/for-authors/latex-issues/latex-examples
\usepackage[varg]{txfonts}

\usepackage{mathrsfs}
\usepackage{mathtools}
\usepackage{amstext}
\usepackage{hyperref}
\usepackage[none]{hyphenat}
\usepackage{color}

\usepackage{amsmath,amssymb}

\begin{document}
\title{SIV Guided Resolution of the $^7$Li BBN Problem}

%\shorttitle{$^7$Li BBN Problem within SIV \& RISS}
%SIV affected departure from TD equilibrium

%$^{1}${Institute for Advanced Physical Studies, Sofia, Bulgaria}\\%111 “Tsarigradsko shose” Blvd, Sofia 1784
%$^{2}${Ronin Institute for Independent Scholarship, Montclair, NJ, USA}\\ %127 Haddon Pl.
%$^{3}${National Coalition of Independent Scholars,  Battleboro, VT, USA}\\ %125 Putney Rd, Battleboro, VT 05301

\author{V. G. Gueorguiev}
\institute{Institute for Advanced Physical Studies, Sofia, Bulgaria
\and National Coalition of Independent Scholars,  Battleboro, VT, USA
\and Ronin Institute for Independent Scholarship 2.0, Sacramento, CA, USA}

\date{Received July 11/ Revised Aug. 27/ Accepted Sep. 10, 2025}

%\begin{abstract}
\abstract{
A possible resolution of the $^7$Li problem within the Standard Model Big-Bang Nucleosynthesis is presented.
The key idea originates from the application of the Scale-Invariant Vacuum (SIV) paradigm to the BBN.
However, here we arrive at the conclusion  that Reparametrization Invariant Symmetry Scaling (RISS) 
is the more appropriate framework for the epoch of the BBN and use the SIV only as a guidance framework.
The outcome is $\chi^2<0.04$ fit to the observed primordial abundances of $^4$He, D/H, $^3$He/D, 
and fit of $\chi^2\approx1$ when including $^7$Li/H observations.
The results are obtained and compared to the known standard BBN values 
by utilizing the publicly available PRIMAT code. 
The resolution of the  $^7$Li problem requires SIV-guided deviation from the local thermal equilibrium during BBN,
such that the thermal energy of matter and radiation scale differently 
with respect to the SIV-conformal factor $\lambda$ during the BBN epoch.
This may be viewed as conformal symmetry breaking due to cooling of plasma and the properties of matter.
As such, the framework may be of relevance to the problem of the nuclear fusion  as well.
The deduced baryon matter content is $\Omega_b\approx12\%$ for unbroken SIV and 
$\Omega_b\approx38\%$ for partially broken SIV, 
but with $\lambda<1$ in both cases, which signals preference for 
Reparametrization Invariant Symmetry Scaling (RISS) over the conventional SIV viewpoint.
Applying the RISS paradigm results in  $\lambda>1$ and $\Omega_b\approx10\%$ with 
clear departure of $n_T$ away from the naive SIV suggested value.
In all the cases where the $^7$Li problem is resolved, 
the baryon content is significantly higher than the usually accepted value 
of $\Omega_b\approx4.9\%$  within the $\Lambda$CDM.
}%\end{abstract}
\maketitle

%VGG 
\nolinenumbers % deactivates

\section{Introduction and Background Framework}\label{sec:intro}

Big Bang nucleosynthesis (BBN), first proposed by \cite{Gamow'46} and later studied in detail by \citet{Coc'17, Iliadis'20}, predicts the primordial abundances of light elements such as deuterium and helium-4 \citep{Wagoner'67, Steigman'07, Coc'17, Iliadis'20}, in close agreement with astrophysical observations. This match is a cornerstone of the standard cosmological model \cite{Coc'17}. A major exception is lithium-7: standard BBN overestimates its primordial abundance by a factor of three to four compared to observations of metal-poor halo stars \citep{Cyburt'03, Coc'04, Cuoco'04, Asplund'06, Sbordone'10, Hou'17, Iliadis'20}. This discrepancy, known as the \textit{Cosmological Lithium Problem} \citep{Boesgaard'85, Steigman'07, Hou'17}, has persisted despite decades of attempts involving astrophysical, nuclear, and particle-physics solutions \citep{Fields'11, Cyburt'16, Coc'17, Iliadis'20}. Its persistence shows limits in our knowledge of stellar astrophysics and nuclear processes, and may even suggest new physics beyond the Standard Model \citep{Hou'17, Coc'17}. Solving this problem is crucial to test the strength of BBN and probe for new fundamental physics.

The \textit{Scale-Invariant Vacuum} (SIV) paradigm provides analytic expressions that allow an initial study of BBN \citep{Maeder19,VG&AM'23}. 
This capability makes the SIV framework promising for BBN studies with computational tools such as the PRIMAT Mathematica code \citep{PRIMAT}.
The SIV paradigm has explained several astrophysical phenomena, including galactic rotation curves and density fluctuations, without dark matter \cite{MaedGueor20a,SIV-PartII'24}. Despite its Lagrangian formulation \cite{SIV-PartI'23}, the framework lacks a dynamical basis for $\lambda$. Instead, it relies on heuristic arguments such as scale invariance of empty space and large-scale homogeneity, which may not hold at all scales or in matter-dominated regions. To address this, a new framework—\textit{Reparametrization Invariant Symmetry Scaling} (RISS) \cite{RISSandDE'24, Gueorguiev'25}—has been proposed. Here, the gauge factor arises from time reparametrization invariance of physical laws rather than from vacuum symmetry assumptions \cite{GueorMaed25}. 
The RISS paradigm, while rooted in scale invariance, treats the conformal scale transformation differently and explores its implications for physical laws \citep{RepInv-and-KeyProps-of-Phys-Sys'21}. This leads to distinct predictions for processes such as primordial nucleosynthesis, the focus of this work. We extend the SIV approach to BBN beyond the analytic results of \citep{Maeder19} and \citep{GueorMaed25}, by applying it within the \texttt{PRIMAT} code \citep{PRIMAT}.

Early Universe cosmology, essential for BBN, rests on fundamental constants. These include $H_0$, $\rho_{c0}$, $G$, $T_0$, the black-body constant $a_{BB}$, and Boltzmann’s constant $k_B$. In computational tools such as PRIMAT \cite{PRIMAT}, natural units are used ($c=1$, $k_B=1$, $\hbar=1$) with cosmological parameters $\Omega_m=0.31$ and $\Omega_b=0.05$. Modeling BBN requires solving thermonuclear reactions that govern nuclear abundances, expressed as $Y_i = n_i / n_b$. 
Reaction rates are $\Gamma_{ij\rightarrow{kl}}=n_b\gamma_{ij\rightarrow{kl}}$, in units of s$^{-1}$. Forward and reverse rates satisfy 
$\bar{\gamma}_{j\ldots\rightarrow{i\ldots}} = \gamma_{i\ldots\rightarrow{j\ldots}} = \gamma_{j\ldots\rightarrow{i\ldots}} \times \alpha T_9^\beta \exp(\gamma/T_9)$, with $T_9$ in GK, assuming local thermodynamic equilibrium. Time evolution follows the Friedmann-Lemaître-Robertson-Walker equation, $\dot{a}/a = H = \sqrt{8\pi G \rho / 3}$, which integrates to the cosmic time variable $\tau$. 
The total energy density $\rho$ includes radiation, matter, and corrections from neutrino decoupling. The latter introduces a distortion factor $\mathscr{\mathcal{S}}(T)$ into the temperature–scale factor relation: 
$T a(T)\mathscr{\mathcal{S}}^{1/3}(T)=T_0 a_0$, relevant near electron-positron annihilation. 
The $\mathscr{\mathcal{S}}(T)$ affects the baryon-to-photon ratio $\eta$, linking constants, energy densities, time evolution, and the thermal history of the  Universe \cite{VG&AM'23}.

\section{SIV Guided Method} \label{sec:Method}

The SIV framework for BBN introduces a near-constant scaling parameter $\lambda$, evolving from $1/t_{\rm in}$ in the early Universe to unity today. A key distinction is the use of different \textit{scaling exponents for rest-mass energy} ($n_m = +1$) and thermal energy ($n_T = -1/2$); that is, $m\rightarrow m\lambda^{n_m}$ and $T\rightarrow T\lambda^{n_T}$. This follows from the conservation laws for matter ($w=0$) and radiation ($w=1/3$), governed by the SIV relation $\rho_w a^{3(1+w)} \lambda^{1+3w} = \rho_0$ \cite{Maeder17a,Maeder19}. Despite their difference, energy conservation holds \cite{VG&AM'23} since the $\lambda$-scaling of radiation energy matches rest-mass energy scaling in finite systems during photon absorption or emission. 
This imposes the constraint $n_m - 3 = 4n_T$, consistent with the SIV values.

These scalings require changes to thermonuclear reaction rates in tools like PRIMAT. 
Forward rates are modified by a temperature factor $\lambda^{n_T}$, 
while reverse rates include combined mass and temperature scalings, $\lambda^{n_m+n_T}$ and $\lambda^{n_m-n_T}$, 
for terms such as $\gamma/T_9$. The time parametrization in the SIV frame, $d\tau' = \lambda d\tau$, means the Einstein General Relativity (EGR) frame rates must also be rescaled by $\lambda$. This agrees with the $\lambda$-scaling of $G\rho$ from the FLRW equation and ensures consistency across epochs. Radiation and matter energy densities are modified by $\lambda^{4n_T}$ and $\lambda^{n_m-3}$, respectively, maintaining $\rho$ scaling and time evolution. The PRIMAT scale factor $a(T)$ and functions such as $\delta\rho(T)$ are corrected by rescaling their temperature arguments by $\lambda^{n_T}$, giving a self-consistent SIV integration into BBN computations. The fit quality parameter $\chi^2$ is then used to compare theory with observations, scanning parameter space for $\Omega_b$ and $\lambda$ while fixing $\Omega_d=0$, since dark matter has negligible impact in the radiation-dominated BBN epoch.

This gives the first modifications for forward and reverse reaction terms:
$%\begin{eqnarray*}
{\gamma}_{j\ldots\rightarrow{i\ldots}}(T)\rightarrow
{\gamma}_{j\ldots\rightarrow{i\ldots}}(\lambda^{n_T}T),
$
$
\alpha T_9^\beta \exp{\left(\gamma/T_9\right)}\rightarrow
\alpha (\lambda^{n_m+n_T}T_9)^\beta \exp{\left(\lambda^{n_m-n_T}\gamma/T_9\right)}. 
$ %\end{eqnarray*}
The scaling factors used throughout the text are:
%\begin{itemize}
%\item 
\v{T} = $\lambda^{n_T}$ – overall temperature rescaling for forward reactions;
%\item
\(\text{m\v{T}} = \lambda^{n_m + n_T}\) – combined scaling for mass and temperature;
%\item 
\(\text{Q/\v{T}} = \lambda^{n_m - n_T}\) – scaling for energy-to-temperature ratios such as \(\gamma / T_9\) in reverse rates.
%\end{itemize}
Auxiliary functions like $\delta\rho(T)$ and $\mathcal{S}(T)$ also require modification. 
These are obtained by pulling back the corresponding EGR functions. 
For example, the SIV version of $\mathscr{\mathcal{S}}'(T')$ is 
$\mathscr{\mathcal{S}}(T) = \mathscr{\mathcal{S}}'(T'(T)) = \mathscr{\mathcal{S}}'(T\lambda^{n_T})$, 
and similar transformations apply to dimensionless functions like $\delta\rho(T)$.

\section{Results}\label{sec:Results}

The results are summarized in Table~\ref{Table1}, showing the abundances of key elements produced during BBN. The second column lists observed values, and the third shows results from the PRIMAT code. For this work, PRIMAT was run with a reduced network of thirteen reactions, neutrinos treated as decoupled, and QED dipole corrections omitted. These corrections have only a minor effect and are negligible for this study.

%### Observational and Computational Concordance
{\bf Baseline PRIMAT Predictions:} Using PRIMAT with standard cosmological parameters, \(\Omega_b = 4.9\%\) for baryons and \(\Omega_m = 31\%\) for total matter, gives abundances consistent with observations. Setting \(\Omega_d = 0\) (so \(\Omega_m = \Omega_b\)) leaves the results unchanged. This reflects the radiation-dominated epoch, where nucleosynthesis depends on radiation density and baryon-photon interactions, not dark matter.

{\bf Marginal Impact of Neutrino and QED Effects:} 
Partially decoupled neutrinos or QED dipole corrections cause only small shifts and do not affect the main conclusions.
Standard-model neutrino decoupling and finite-temperature/QED effects 
produce shifts of order $10^{-3}$–$10^{-4}$ in BBN yields, leaving the overall pattern unchanged \citep{Froustey'20,PRIMAT, Pitrou'20}. 
This confirms the simplified model is sufficient to capture the main SIV features.

{\bf Dark Matter Irrelevance in the BBN Context:} As expected, dark matter plays no role during the radiation-dominated epoch. Nucleosynthesis processes are unaffected by its presence or absence, showing that radiation sets the elemental abundances.

%### Search for Optimal \(\chi\):  
A parameter-space search reveals minimal \(\chi\) values for $\lambda > 0.5$ with $\Omega_d = 0$. The scaling exponents $n_T$ and $n_m$ were varied to test their impact on abundances. 
Earlier SIV-guided methods \cite{VG&AM'23} could not fully probe regions near $\lambda = 1$, where the lowest \(\chi\) often occurs for $n_T = n_m$. To address this, a direct search was used, avoiding constraints on initial conditions and extending the analysis into unexplored regimes. Fits were performed over $\Omega_b$, $\lambda$, and the exponents $n_T$ and $n_m$, treated as real numbers.   

%### Reflection on Table~\ref{Table1}:  
Table~\ref{Table1} highlights the robustness of the BBN formalism without dark matter. The results show not just numerical consistency but also theoretical coherence within SIV, even where it departs from standard approaches. The agreement with observed abundances, combined with the novel $\lambda$-based scalings, points to the promise of SIV as a cosmological framework.

\begin{table*}%[htb]
\tiny
\begin{center}
\begin{tabular}{|c|cc|cc|ccc||cc||cc|cc||c||}
\hline\hline 
Element & Obs. & PRMT &  SIV$_0$ & $\hbar$ & rad & mat & $\text{fit}_0$ & SIV$_1$ & {\bf fit$_1$} & $\lambda=1$ & {\bf fit$_2$} & SIV$_{-1}$ & fit$_{-1}$ & {\bf fit$^{*}$}\\
\hline 
$\Omega_{b}\; ~ [\%]$ & 4.9 & 4.9 & 3.6 & 4.5 & 4.2 & 4.8 & 20. & 12. & 38. & 38. & 18. & 3.0 & 7.5 & 10. \\
\hline 
 $\lambda$  & -- & -- & 1.06 & 1.01 & 1.03 & 0.99 & 0.66 & 0.89 & 0.77 & 1 & 1 & 1.2 & 2.3 & 1.2 \\
 $n_T$ & -- & -- & $-\frac{1}{2}$ & -1 & $-\frac{1}{2}$ & 1 & -0.29 & $-\frac{1}{2}$ & -0.55 & 0 & 0 & $-\frac{1}{2}$ & 0.02 & 0.9 \\
 $n_m$ & -- & -- & 0 & -1 & $-\frac{1}{2}$ & 1 & -0.29 & 1 & 0.8 & 0 & 0 & -1 & -1 & -1 \\
\hline
 \text{\v{T}} & -- & -- & 0.97 & 0.99 & 0.98 & 0.995 & 1.13 & 1.06 & 1.15 & 1.15 & 1.15 & 0.92 & 1.02 & 1.17 \\
 \text{m\v{T}} & -- & -- & 0.97 & 0.98 & 0.97 & 0.99 & 1.28 & 0.95 & 0.94 & 0.94 & 0.91 & 0.79 & 0.43 & 0.98 \\
 \text{Q/\v{T}} & -- & -- & 1.03 & 1. & 1. & 1. & 1. & 0.85 & 0.71 & 0.71 & 0.7 & 0.92 & 0.42 & 0.72 \\
\hline 
 \text{$\eta_{10}$}& 6.09 & 6.14 & 4.47 & 5.69 & 5.24 & 5.97 & 25.1 & 14.8 & 48.3 & 47.6 & 22.7 & 3.8 & 9.45 & 12.5 \\
\hline \hline
 \text{H} & 0.755 & 0.753 & 0.755 & 0.755 & 0.755 & 0.755 & 0.755 & 0.756 & 0.755 & 0.741 & 0.756 & 0.767 & 0.732 & 0.755 \\
 $Y_P\text{=4}Y_{\text{He}}$ & 0.245 & 0.247 & 0.245 & 0.245 & 0.245 & 0.245 & 0.245 & 0.244 & 0.245 & 0.259 & 0.244 & 0.235 & 0.268 & 0.245 \\
 $\text{D/H}\times10^{5}$& 2.53 & 2.43 & 2.53 & 2.53 & 2.53 & 2.53 & 2.53 & 2.53 & 2.52 & 0.748 & 2.53 & 2.60 & 2.52 & 2.53 \\
\hline 
 $^{3}\text{He/H}\times10^{5}$ & 1.1 & 1.04 & 1.01 & 1.05 & 1.04 & 1.05 & 1.15 & 1.23 & 1.50 & 1.03 & 1.51 & 1.06 & 1.84 & 1.5 \\
 $^{7}\text{Li/H}\times10^{10}$ & 1.58 & 5.56 & 5.85 & 5.34 & 5.43 & 5.29 & 3.97 & 3.31 & 1.79 & 5.4 & 1.77 & 4.81 & 0.83 & 1.8 \\
 \hline 
  $\chi$ & -- & 6.84 & 7.12 & 6.27 & 6.42 & 6.19 & 3.99 & 2.91 & {\bf 1.07} & 30.4 & {\bf 1.07} & 5.73 & 3.63 & {\bf 1.06} \\
% $N_{\text{eff}}$  & 3.01 & 3.01 & 3.01 & 3.01 & 3.01 & 3.01 & 3.01 & 3.01 & 3.01 & 3.01 & 3.01 & 3.01 & 3.01 & 3.01 \\
% $\Omega _m$ & 0.3097 & 0.3097 & 0.2963 & 0.306 & 0.3024 & 0.3083 & 0.461 & 0.1182 & 0.3855 & 0.3797 & 0.181 & 0.6175 & 0.07761 & 0.5864 \\
% \text{$\chi $D}  & 1. & 2.308 & 0 & 0 & 0 & 0 & 0.005968 & 0.09236 & 0.1134 & 42.01 & 0.1418 & 2.778 & 4.053 & 0.07986 \\
% \text{$\chi $He3} & 1. & 1.894 & 0.2475 & 0.1589 & 0.1747 & 0.1496 & 0.1565 & 0.3802 & 1.161 & 34.3 & 1.176 & 2.271 & 3.942 & 1.15 \\
\hline\hline 
\end{tabular}
\caption{\label{Table1} \tiny
Abundances of light elements in standard and SIV-guided BBN. 
Observational uncertainties are 1.6\% for $Y_{P}$, 1.2\% for D/H, 18\% for He/H, and 19\% for Li/H. 
The first four columns reproduce results from the initial SIV study (2023) and the 2025 update. %\cite{VG&AM'23}. 
The next four columns give best-fit values of $\Omega_b$ and $\lambda$, enforcing local thermodynamic equilibrium with $n_T=n_m$, for specific cases discussed in the text. The following two columns show results for the SIV-motivated choice $n_T=-1/2$, $n_m=1$, and for the corresponding fit over $n_T$ with the constraint $n_m=3+4n_T$. The $\lambda=1$ case (analogous to Tsallis non-extensive statistics) illustrates the failure of standard time determination with $\lambda=1$, while keeping the nuclear-reaction scaling factors \v{T}, m\v{T}, and Q/\v{T} from the fit$_1$ case. The next column presents the best fit over $\Omega_b$, temperature, and mass scales encoded in \v{T}, m\v{T}, and Q/\v{T}, with $\lambda=1$ and standard time determination. The last three columns explore the scenario with unaltered $\hbar$ by fixing $n_m=-1$, while allowing $n_T$ to vary away from $-1/2$, so that the scaling of $G\rho$ comes only from $G$. The final column corresponds to $n_\rho=-2.75$ for $G\rho$ scaling via $\lambda^{1-n_m+n_\rho}$, unlike previous cases where it was set to zero because $\rho$-scaling was induced by temperature rescaling.
}
\end{center}
\end{table*}

The fourth to seventh columns show best-fit results for a two-parameter fit over $\Omega_b$ and $\lambda>0.5$, 
for specific choices of $n_T$ and $n_m$. 
The first case (column four, SIV$_0$) follows the values motivated by earlier studies of BBN in the SIV framework \cite{VG&AM'23}. 
Here, $n_T=-1/2$ and $n_m=0$, reflecting the expectation that in the radiation epoch $n_T=-1/2$, 
and that the pull-back of relevant functions may also allow $n_m=0$, 
as discussed in \cite{VG&AM'23}.

In the following cases, radiation and matter are in local thermal equilibrium, so both scale the same way ($n_T=n_m$). 
Specific values can then be assigned. 
The fifth column, labeled $\hbar$, reflects the assumption that $\hbar$ acts as a conversion constant. 
Since thermal energy $E=k_BT$ is linked to the Planck relation $E=\hbar\omega$, 
the SIV rescaling of time intervals $d\tau'=\lambda d\tau$ implies $k_BT' = k_BT\lambda^{-1}$; 
thus, $\hbar$ behaves as an in-scalar in this case. 
The next columns further explore the condition $n_T=n_m$ for SIV-motivated epochs. 
Table~\ref{Table1} shows that these cases resemble the standard PRIMAT results, 
but with slightly different $\Omega_b$ values. 
Imposing $n_T=n_m$ tends to drive $\lambda$ toward 1, 
in sharp contrast with the SIV expectation that $\lambda$ should be greater than 1. 
In fact, values of $\lambda<1$ become prominent in the next set of columns. 
The eighth column (fit$_0$) represents a three-parameter fit over $-1<n_T<1$, 
together with $\Omega_b$ and $\lambda$, but with $n_T=n_m$.

For all these cases the goodness of fit indicator $\chi$ is very small (almost zero or less than 0.1) when considering only $^4$He and deuterium data,
when $^3$He is added $\chi$ is between 0.15 and 0.25 indicating that the $^7$Li is the hardest data point to be described with such model parameters.
The situation changes in the subsequent cases where  $n_T\ne\,n_m$ that brings the over all $\chi$ below 6 due to improvement on the $^7$Li.
In these cases $\chi$ for $^4$He and deuterium is still often less than 0.1, but when $^3$He is added then $\chi$ could be above 1 showing that 
$^3$He is becoming a challenging data point as well.

The SIV-guided deviation from local thermal equilibrium begins with the two-parameter fit over $\Omega_b$ and $\lambda$, shown in column SIV$_1$. 
Column fit$_1$ is a three-parameter fit over $-1<n_T<1$, $\Omega_b$, and $\lambda>0.5$, similar to fit$_0$ but with the constraint $n_m=3+4n_T$. 
The first column with $\lambda=1$ is not a fit; it illustrates the effect of reverting to the standard time determination while keeping the rescaling factors \v{T}, m\v{T}, and Q/\v{T} from the fit$_1$ case. The next column shows a three-parameter fit over $\Omega_b$ with temperature and mass scalings restricted to within 20\% of the fit$_1$ values. These results highlight the importance of distinct temperature and mass scalings for addressing the $^7$Li problem. 
While fit$_2$ is a minimalistic fit, and loosely resembles Tsallis non-extensive statistics \citep{Hou'17}, it is difficult to justify in a simple way, as discussed in the next section.  

The last three columns adopt the RISS perspective, where $\hbar$ is fixed and $n_m=-1$. 
The first choice keeps the SIV expectation $n_T=-1/2$, which does not solve the $^7$Li problem. 
This may reflect violation of the conserved quantity $\rho_w a^{3(1+w)}\lambda^{1+3w}=\rho_{0}$ and departures from $n_T=-1/2$ and $n_\rho=-2$, 
as suggested by the role of the distortion term $\mathcal{S}(T)$. 
Such non-conservation also alters the scaling of $\rho$. 
One must then consider $G\lambda^{3-n_m-2n_T+n_\rho}\rho(T\lambda^{n_T})$, 
which reduces to $G\rho(T)\lambda^{4n_T}$ when $n_m=-1$, $n_\rho=-2$, and $n_T=1$. 
There may also be an extra factor $\lambda^{n_\rho}$, as explained in the caption of Table~\ref{Table1}. 
Notably, for $w=1/3$, $n_T$ is not $-1/2$ but closer to 0.9.
\begin{figure}[h]
\centering
\includegraphics[width=0.9\columnwidth]{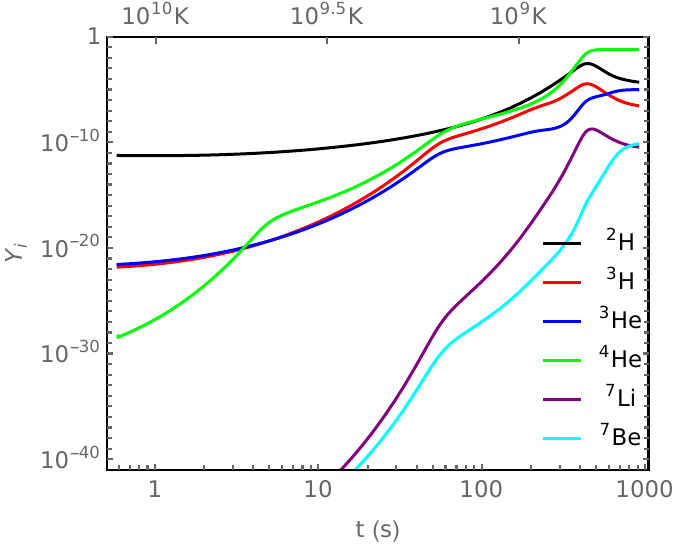}
\caption{\label{abundances}
Element abundance for the case corresponding to fit$^*$ in Table \ref{Table1}.
Notice the similarity with the figures in the Appendix of \cite{VG&AM'23} 
for the standard BBN and the SIV case discussed there.}
\end{figure}

\section{Interpretation and Conclusions}\label{sec:Summary}

The SIV paradigm introduces specific modifications to reaction rates and their temperature dependence, 
necessary to maintain consistency between the EGR and WIG frames. 
As Table \ref{Table1} shows, some cases produce $\Lambda$CDM-compatible baryon densities around $\Omega_b\approx 4\%$, 
but most do not solve the $^7$Li problem. 
Three cases yield $\chi\approx1$, which usually indicates agreement between theory and observations within uncertainties. 
These cases offer a potential solution to the $^7$Li problem, which could improve further if astrophysical processes reduce $^7$Li abundances \citep{2024PhRvD.110c0001N}. 
Pure astrophysical effects cannot reduce $^7$Li by a factor of 3 to 4, 
but along with possible $^3$He adjustments, it may help bring overall $\chi$ below 1 using the current approach.

The results suggest that resolving the $^7$Li problem may involve a deviation from local thermal equilibrium. 
An increased effective temperature (\v{T} $>1$) is required in all cases with $\chi<6$. 
Within conventional physics, this could reflect local inhomogeneous matter moving with specific velocity, 
experiencing a higher surrounding radiation temperature. 
Even though this aligns with the idea that high-energy particles approach a "massless" state, 
the reduced mass-energy scale (m\v{T} $<1$, Q/\v{T} $<1$) is difficult to justify without a specific model. 
Tsallis non-extensive statistics \citep{Hou'17} also provides no physical explanation for such deviations. 

The $\lambda$-independent fit (fit$_2$) with $\chi\approx1$ yields \v{T}, m\v{T}, and Q/\v{T} values similar to fit$_1$, the SIV-motivated relational choice; 
fit$_1$ indicates partial breaking of scale-invariant symmetry, as $n_T$ deviates from the SIV value $-1/2$. 
Although \v{T}, m\v{T}, and Q/\v{T} are similar for fit$_1$ and fit$_2$, differences in $\Omega_b$ arise from different constructions of the time parameter.

Applying SIV rules requires different scaling for matter and radiation. 
The best fit with $n_T=n_m$ (fit$_0$) cannot reach $\chi\approx1$, 
while the SIV-motivated choice $n_T=-1/2$ with $n_m=3+4n_T$ approaches $\chi\approx1$ at $n_T=-0.55$ (fit$_1$). 
This indicates broken scale invariance due to the presence of matter. 
The distortion term $\mathcal{S}(T)$, arising from annihilation processes, modifies the relationships between $a$, $\rho$, $T$, and $\lambda$, 
implying deviations from conservation of the SIV quantity $\rho_w a^{3(1+w)}\lambda^{1+3w}$ 
and thus departures from the SIV-predicted $n_m$ and $n_T$. 
This departure can be accommodated within the RISS framework.

SIV and RISS both use a time-dependent conformal factor $\lambda(t)$, with conformal transformations keeping $c$ fixed. 
SIV begins with a general $\lambda(x)$ and restricts to time dependence due to large-scale homogeneity and isotropy. 
Neglecting $\hbar$ and keeping $G$ constant is reasonable for cosmological units, 
but may fail on smaller scales where homogeneity breaks down. 
RISS, on the other hand, starts with $\lambda(t)$ derived from the fundamental symmetry of 
time reparametrization invariance for the description of process, applicable at all scales. 
Thus, keeping $\hbar$ constant is relevant for BBN within RISS.

SIV breaking is also evident when $\lambda<1$, which conflicts with the expected SIV gauge form $\lambda=t_0/t$, 
where $0\le t_\text{in}<t\le t_0$ implies $\lambda>1$ before the present time $t_0$, often set to 1. 
The $\lambda<1$ issue suggests that reparametrization invariance may provide a more general view than enforcing the SIV gauge. 
While SIV and RISS lead to similar equations \citep{sym13030379}, RISS does not fix the functional form of $\lambda(t)$. 
Then, the specific value of $\lambda$ can be interpreted as a correction to the overall scale of the distortion factor $\mathscr{\mathcal{S}}(T)$ defining $a(T)$. 
This scale correction could reflect the effects of reheating and recombination that establish the observed CMB temperature $T_0$. 
From the PRIMAT expression $T\,a(T) = a_0 T_0 / \mathscr{\mathcal{S}}^{1/3}(T)$, taking $a\rightarrow a_0=1$ gives $T_0$, 
but this is after BBN and does not include reheating and recombination. 
Thus, the SIV $\lambda$-correction to $a(T)$ restores the correct post-BBN temperature, expected to differ from the standard CMB value: 
$T_0 \rightarrow T_\text{end}^\text{BBN} = T_0 \lambda^{-1/2}$. 
In the author's experience, this correction to $\mathscr{\mathcal{S}}(T)$ or $T_0$ alone can resolve the minor discrepancy between theory and observation in deuterium abundance. Potential primordial deuteron tension was indicated by \cite{Pitrou'21}, 
but this arose from differing rate evaluations. 
Comparisons of BBN codes (PRIMAT/PRyMordial/PArthENoPE)
show small numerical differences once networks and rates are harmonized \cite{Giovanetti'24}.
This also suggests future potential test and validations of the current results by using alternative BBN codes and reaction rates.
Since SIV and RISS rely on time-dependent conformal transformations, obtaining a nearly constant $\lambda(t)$ to fit light-element abundances indicates conformal symmetry breaking due to matter production and plasma temperature reduction.

The interpretation of $\lambda$ remains unclear within SIV. 
In RISS, however, it can be viewed as a gravitational effect on the unit of time. 
In the early Universe, when matter was denser, gravitational influence on atomic clocks could shorten the effective time unit, 
which can be represented by $\lambda$. 
If $\lambda$ is constant, this is equivalent to a change of units that does not alter physics. 
If $\lambda(t)$ varies with time, equations must remain reparametrization invariant to avoid artificial effects in physical conclusions \citep{RISSandDE'24,Gueorguiev'25}.

It has been shown that assuming $\lambda\approx\text{const.}$ during BBN allows one to use $\mathscr{\mathcal{S}}(T)$ within the SIV framework, 
or treat SIV effectively as a background effect within standard BBN. 
This produces results compatible with standard BBN (see Table~\ref{Table1}). 
To fit observational data, one must depart from the standard SIV expression $\lambda=1/t_\text{in}\approx1/\Omega_m^{1/3}\ge1$, 
which is linked to the analytical solution $a(t)$ as a function of $\Omega_m$ \citep{Jesus18}. 
This departure allows significant reduction of $\chi$, as seen in columns SIV$_1$ and fit$_1$ of Table~\ref{Table1}. 

A pure SIV calculation with $n_T=-1/2$ and $n_m=1$ gives $\Omega_b\approx12\%$, 
while partially broken SIV yields $n_T=-0.55$, $n_m=0.79$, and $\Omega_m\approx38\%$. 
The higher baryon content also increases the photon-to-baryon ratio to $\eta_{10}\approx15$, 
about 2.5 times the standard value $\eta_{10}=6.09$. 
In this study, $\eta$ was treated as an in-scalar based on baryon-to-photon number ratios, 
but the expression may need reevaluation if temperature is used as a proxy for photon density.

The differences in the specific $\Omega_m$ values should be viewed as artifacts of model assumptions. 
Determining the appropriate model parameters and achieving the same level of concordance as standard BBN with $\Lambda$CDM remains a long-term task. 
Such exploration may also shed light on the Hubble tension. 
The discrepancy between the CMB-inferred and locally measured Hubble constants could potentially be addressed within the SIV gauge, 
since the relevant cosmic evolution is mostly in the matter-dominated epoch. 
However, for the early Universe, the RISS paradigm may be more appropriate. 
Its ultimate test would involve using the inverse temperature as the time parameter for BBN, $t=T_0/T$, 
which would also require extending the FLRW equations to a fully reparametrization-invariant form.

The main conclusion is that the SIV paradigm offers useful guidance for constructing a BBN model 
compatible with the observed abundances of $^4$He, D/H, T/H, and $^7$Li/H, as achieved in standard BBN. 
While SIV faces the same $^7$Li problem as standard BBN, it also suggests a potential SIV-guided departure from local thermal equilibrium, 
as interpreted within RISS, which could help resolve the $^7$Li problem. 
This departure may also be relevant for nuclear fusion processes.

%\section*{Acknowledgments} This research does not receive any specific grant from funding agencies in the public, commercial, or not-for-profit sectors. The author is grateful for the moral and financial support by particularly close private parties during the various stages of the research presented and to Prof. A. Maeder for the long and fruitful scientific collaboration over the years. 

%\section*{Data availability}No new data was generated or analyzed in support of this research.

%% If you have bibdatabase file and want bibtex to generate the bibitems, please use

%\bibliographystyle{elsarticle-num-names} %elsarticle-harv %elsarticle-num-names

%VGG%

%\bibliographystyle{elsarticle-num-names}%elsarticle-names %elsarticle-harv %elsarticle-num-names%plain
%\bibliographystyle{apsrev4-1}

%\bibliographystyle{mnras}
%\bibliography{../../Primordial_Deuterium/Primordial_Deuterium}

%\printbibliography
%\end{document}
% Set the starting number of the bibliography

\bibliographystyle{aa}
%VGG%\bibliography{Primordial_Deuterium}
%\bibliography{../../DM&DEwithinSIV}
%\bibliography{../../Primordial_Deuterium/Primordial_Deuterium}
%\input{SIVguidedBBNS_v12A&A.bbl}

%\printbibliography
%\end{document}

\end{document}